\documentclass[reprint,nofootinbib,amsmath,amssymb,aps,prd]{revtex4-2}

\usepackage{graphicx}
\usepackage{geometry}
\usepackage[english]{babel}
\usepackage{amsmath}
\usepackage{physics}
\usepackage{tensor}
\usepackage{hyperref} 
\usepackage{cleveref}
\usepackage{amsfonts}
\usepackage{comment}
\usepackage{mathtools}
\usepackage[normalem]{ulem}

\usepackage{soul} %for comments
\setstcolor{cyan} %for comments

\usepackage{dcolumn} % Align table columns on decimal point
\usepackage{bm}      % Bold math

\usepackage{xcolor}

\usepackage{orcidlink}
 \newcommand\pd[2]{\frac{\partial#1}{\partial#2}}

\newcommand{\Say}[1]{``#1''}

\usepackage{phfcc}

\begin{document}
% Authors %%%%%%%%%%%%%%%%%
\preprint{APS/123-QED}
\title{On more general Gibbons-Hawking-York boundary terms for \texorpdfstring{\boldsymbol{$f(R)$}}{f(R)} gravity}

\author{Federico Scali\orcidlink{0009-0004-0637-561X}}
\email{fscali@uninsubria.it}
\affiliation{Department of Science and High Technology, University of Insubria, Via Valleggio 11, 22100, Como, Italy}
\affiliation{INFN section Milan, Via Celoria 16, 20133 Milan, Italy}

\author{Sergio Luigi Cacciatori\orcidlink{0000-0002-4167-9123}}
\email{sergio.cacciatori@uninsubria.it}
\affiliation{Department of Science and High Technology, Università dell'Insubria, Via Valleggio 11, Como, 22100, Italy.}
\affiliation{INFN section Milan, Via Celoria 16, 20133 Milan, Italy}
\affiliation{Como Lake Centre for AstroPhysics (CLAP), DiSAT, Università dell’Insubria, Via Valleggio 11,
22100 Como, Italy}

\author{Matteo Galaverni\orcidlink{0000-0002-5247-9733}}
\email[]{matteo.galaverni@gmail.com}
\affiliation{Vatican Observatory, V-00120, Vatican City State}
\affiliation{INAF/OAS Bologna, via Gobetti 101, I-40129 Bologna, Italy}
\affiliation{INFN, Sezione di Bologna, via Irnerio 46, 40126 Bologna, Italy}

\author{Gabriele Gionti, S.J.\orcidlink{0000-0002-0424-0648}}
\email{ggionti@specola.va}
\affiliation{Vatican Observatory, V-00120, Vatican City State}
%\affiliation{Vatican Observatory Research Group, Steward Observatory, The University Of Arizona,
%933 North Cherry Avenue, Tucson, Arizona 85721, USA}
\affiliation{INFN, Laboratori Nazionali di Frascati, Via E. Fermi 40, 00044 Frascati, Italy.}

% Abstract %%%%%%%%%%%%%%%%%%%%%%%%%
\begin{abstract}
More general boundary conditions are introduced - with respect to known literature - which produce a well-posed stationary action principle in $f(R)$ gravity. These conditions encompass the known cases, thus enlarging the space of admissible solutions. The corresponding Gibbons-Hawking-York boundary terms are explicitly computed and hold for any analytic $f(R)$. The perspectives of application concerning the scalar-tensor mapping and the laws of black-hole thermodynamics are outlined in the conclusions. 

\begin{comment}
In the first part of this work, more general boundary conditions are discussed - with respect to known literature - which produce a well posed stationary action principle in $f(R)$ gravity. The corresponding Gibbons-Hawking-York boundary terms are computed in two particular cases, in which the variation of the scalar curvature on the boundary is proportional to the variation of the trace of the extrinsic curvature. The compatibility with Gauss-Codazzi equation of the new conditions is also thoroughly discussed, revealing the need of a dynamical boundary.

In the second part of this work, the correspondence of $f(R)$ gravity with scalar-tensor (ST) theory is discussed in light of the new boundary conditions. It is argued how the latter sustain a well posed action principle only in a generalized scenario, in which the ST scalar is a function of the metric and its higher derivatives. In this case, however, it is argued how the $f(R)$ to ST ceases to hold. 

Eventually, the implications concerning the Jordan-to-Einstein frame mapping of ST theory are addressed, and the perspectives in the direction of black holes' thermodynamics are outlined.  
\end{comment}

\end{abstract}
\maketitle

\tableofcontents

\section{Introduction}
\subsection{History}

Already in 1912 Einstein had understood and developed many of the essential ingredients of general relativity (GR); namely, the basic mathematical objects describing the gravitational field as a geometric property of spacetime and the motion therein of matter particles and of light \cite{Einstein1911}. From that moment, Einstein struggled for three years to formulate the correct field equations for the new theory, stuck on the riddle of general covariance and its physical implications. At that time, Hilbert was mainly dedicated to problems in mathematical physics \cite{Sauer1998} - particularly concerning the application of his theory of linear integral equations \cite{Hilbert1912integralgleichungen} - under the influence of the so-called \Say{electromagnetic world view} prevailing in Göttingen at the beginning of the 20th century.\footnote{This generated attempts like Mie's theory \cite{Mie1912-1913grundlagen}, in which the mass of the electron is tentatively explained through the electromagnetic interaction.} Therefore, when in the summer of 1915 Einstein delivered a six lectures course at the University of Göttingen on his general theory of relativity, Hilbert was immediately drawn to the idea of a unified theory of the gravitational and the electromagnetic fields within the new mathematical framework, regulated by generally covariant field equations.\footnote{This was also triggered by Mie's recent unified theory for the structure of the electron \cite{Mie1912-1913grundlagen,Sauer1998} and the attempts of Born for its covariant reformulation.} Events took a turn in November of the same year, when in a correspondence with Hilbert, Einstein informed the mathematician about its return to generally covariant field equations, after the experience of the noncovariant \textit{Entwurf} theory developed with Grossmann two years earlier \cite{Einstein_Grossmann1913entwurf}. Alarmed, Hilbert sped up his work and on the 20th of November submitted his generally covariant \textit{fundamental equations of physics} for the gravitational and the electromagnetic fields, derived from a stationary action principle \cite{Hilbert1915}. Five days later, Einstein submitted to the Prussian academy of Berlin his \textit{fundamental equations of gravitation} \cite{Einstein1915Equations}, containing equivalent equations for gravity and matter.\footnote{The \textit{priority dispute} between Hilbert and Einstein in formulating the field equations of GR is still debated among historians \cite{Earman_Glymour1978einstein}. That said, there seems to be consensus in considering what happened an extreme case of parallel research.} The purely gravitational part of the action introduced by Hilbert in \cite{Hilbert1915}, involving the Ricci scalar, would have gone down in history as the Einstein-Hilbert (EH) action. See \cite{damour2006once} for a historical review.

The first peculiarity of the EH action is that it involves second order derivatives of the metric field. As a rule of thumb, when all the integrations by parts are settled, variation of an action involving higher than first derivatives in the fields is expected to produce boundary terms depending on the variations of first or higher derivatives in those fields. These terms depend on how the fields change \textit{on and across} the boundary, so that they are not simply determined by specifying proper fields' configurations on the boundary. As a result, the action is not trivially stationary on the solutions of the fields' equations, as it happens for first order Lagrangians, and a finer analysis is needed.

Legitimately, this nuisance did not refrain Hilbert from formulating the action principle for gravity, with the more or less implicit assumption that all the derivatives of the metric would vanish on the boundary sufficiently fast. The problem remained essentially quiescent for almost six decades,\footnote{Einstein's subsequent formulation of an action principle for gravity in 1916 is not an exception, even though he used an action involving a first order Lagrangian: the $\Gamma-\Gamma$ action \cite{Einstein1916Hamiltonsches}. In the paper, he derives the first order Lagrangian by integrating by parts the EH term and imposing $\delta g_{\mu\nu} = 0$ and $\partial_\sigma \delta g_{\mu\nu} = 0$ on the boundary.} until York \cite{York1972} in 1972 referred explicitly to a \Say{true} action for gravity, which, evaluated on vacuum solutions, becomes a surface term involving the trace of the extrinsic curvature of the boundary. The contemporary formulation of the problem, together with its final solution, had to wait the work of Gibbons and Hawking \cite{Gibbons1977} on the path-integral quantization of gravity, a few years later: in order for the gravitational action to be stationary under metric variations that vanish on the boundary - but that may have non-vanishing normal derivatives - the EH action must be endowed with a surface term which is proportional to the trace of the extrinsic curvature of the boundary. Surface contributions added to an action with the sole purpose of cancelling boundary variations would then be called Gibbons-Hawking-York (GHY) boundary terms.

\subsection{Introduction}
The issue of boundary terms, which here was historically introduced in the context of GR, is in fact common to any field theory: given a point in the space of all well-posed boundary conditions,\footnote{A set of boundary conditions is well-posed if it selects a unique solution of the field equations.} the choice of boundary terms will determine whether the action is stationary on the corresponding solution of the Euler-Lagrange equations. If the action is stationary, the variational principle is said to be well-posed. For any choice of boundary term, in general there will be a subspace of the well-posed boundary conditions for which the action is nonetheless stationary. However, restricting to such a subspace also restricts the space of admissible solutions, which should be avoided unless physically motivated. In fact, in concrete physical applications the full space of well-posed boundary conditions is typically not known a priori. The practical challenge is thus to find the boundary term that makes the action stationary on the largest known space of well-posed boundary conditions.

All this discussion about boundary terms may seem formal - after all, if the Euler-Lagrange equations are available, why insist on a stationary action? - until it is clarified why a stationary action is needed. As argued by Dyer and Hinterbichler \cite{Dyer2008}, at the classical level the stationarity of the action is necessary for a well-defined Hamilton-Jacobi theory \cite{Goldstein2002classical} of mechanical systems.\footnote{The Hamilton principal function can be defined as a functional of end point data, whose variation vanishes. Therefore, the Hamilton principal function defines an action for the field theory which is stationary by construction \cite{Goldstein2002classical}.} At the quantum level, a stationary action is required for  the path integral quantization \cite{weinberg1995}. The latter is arguably the most compelling reason: if the action had no stationary points, all the field configurations would be equally suppressed in the classical limit and no classical solution could emerge. That said, the choice of boundary terms typically has some direct physical implications. In GR, for example, the GHY term determines the definition of energy and angular momentum of the gravitational field for asymptotically flat spacetimes \cite{Poisson2009,Arnowitt1959dynamical}. Ultimately, boundary terms affect the laws of black hole thermodynamics \cite{Poisson2009} when these are derived using the Euclidean approach \cite{Dyer2008}. 

At an even deeper level, the question of boundary terms is related to the problem of identifying the degrees of freedom in a field theory. The reason is that, if the action principle is enforced, the available space of well-posed boundary data is determined by the choice of boundary terms. The number of degrees of freedom in a field theory can be defined as the dimension of this space \cite{Dyer2008}; therefore, as argued by Belenchia, Letizia, Liberati and Di Casola in \cite{Belenchia2018higher}, boundary terms can be used as a \Say{diagnostic tool} to identify the physical degrees of freedom in a field theory.\footnote{For example, the addition of the GHY term to the Einstein-Hilbert action confirms that GR propagates only two gravitational degrees of freedom: once this term is included, fixing the configuration of the metric — modulo gauge invariance and constraint equations — is both necessary and sufficient for a well-posed stationary action principle.}

This last point is particularly important for those theories in which the degrees of freedom do not explicitly match the dynamical fields, as it happens for the so-called higher-derivative gravity theories \cite{Saridakis2021ModifiedGravity,Clifton2012ModifiedGravity, Capozziello2010, DeLaurenties2011ExtendedTheoriesGravity, DeLaurenties2009BirdEyeView, Amendola_Tsujikawa2010, DeFelice2010, Sotiriou2010, Nojiri2011, Nojiri2017, Smolic:2013gz}. These are characterized by field equations that are higher than second order in the derivatives of the metric, and are usually obtained by extension of the Einstein-Hilbert action to a functional of higher-order curvature invariants. In these cases, fixing the metric on the boundary is no longer sufficient to ensure a well-posed stationary action principle, and additional boundary conditions must be specified. A common strategy is to remap the higher derivative theory into a dynamically equivalent\footnote{That is, presenting the same field equations of the original theory when the additional fields are on-shell.} second-order theory involving additional dynamical fields, essentially using generalized versions of the Ostrogradsky reduction \cite{Woodard2007, Khodabakhshi2018}. Whenever this is possible, the further required boundary conditions are obtained by fixing the boundary configurations of the additional fields.

In this work, the stationary action principle and the problem of boundary terms are discussed in the context of $f(R)$ gravity \cite{Guarnizo:2010xr}, arguably the simplest higher derivative extension of GR, in which the Einstein-Hilbert Lagrangian is generalized to an arbitrary function of the scalar curvature. As will be shown, the nonlinearity of the $f$ prevents the action principle from being well-posed when only the metric is fixed on the boundary, signaling the presence of additional gravitational degrees of freedom. For extensive reviews of $f(R)$ gravity see \cite{Saridakis2021ModifiedGravity,Clifton2012ModifiedGravity, Capozziello2010, DeLaurenties2011ExtendedTheoriesGravity, DeLaurenties2009BirdEyeView, Amendola_Tsujikawa2010, DeFelice2010, Sotiriou2010, Nojiri2011, Nojiri2017}.

A thorough discussion on the topic was conducted by Dyer and Hinterbichler in \cite{Dyer2008}. The authors motivate the common choice of additionally fixing the scalar curvature on the boundary, $\delta R|_{\mathcal{\partial M}} = 0$, which is inherited from the Ostrogradsky reduction of $f(R)$ gravity to a scalar-tensor theory, essentially by the fact that the corresponding GHY term provides the same Schwarzschild black hole entropy formula as obtained using Wald's Noether-charge method\footnote{Wald's Noether-charge approach for computing the black hole entropy in a metric theory with diffeomorphism-invariant Lagrangian does not rely on an action formulation 
%{and is therefore insensitive to the choice of boundary terms} 
\cite{Wald1993black, Iyer1994some}.} \cite{Wald1993black, Iyer1994some}, as well as the expected ADM energy. In this work, while acknowledging the simplicity of the usual choice, a different perspective is adopted, and the existence of more general boundary conditions compatible with a well-posed stationary action principle is investigated. The central motivation is that a natural mapping of $f(R)$ gravity to a scalar-tensor theory is not, by itself, a physical requirement\footnote{This point was also touched in \cite{Scali2024, Scali2026InductiveApproach}, focusing on non analytic $f(R)$'s.} and cannot justify a restriction of the space of solutions in $f(R)$ gravity. 

More specifically, inspired by the work of Madsen and Barrow \cite{Madsen1989}, a so-called \Say{power-law} boundary condition on the scalar curvature is considered, $\delta R|_{\mathcal{\partial M}} = \alpha K^r \delta K$, where $K$ is the trace of the extrinsic curvature of the boundary, $\alpha$ is a real function of the boundary coordinates and $r$ a real parameter. In this case, which includes both the usual choice, $\delta R|_{\mathcal{\partial M}} = 0$, and Madsen and Barrow result \cite{Madsen1989}, the corresponding GHY term is computed, thereby ensuring a well-posed stationary action principle. The power-law condition is then further generalized to an arbitrary functional condition, $\delta R|_{\mathcal{\partial M}} = g(K) \delta K$, where $g$ is everywhere an analytical function. Even in this case the GHY term is computed and ensures a well-posed action principle. The perspectives for the physical applications of the new GHY terms are discussed in the conclusions. 

\begin{comment}
In the case of the power-law condition, it is shown that the usual mapping of $f(R)$ gravity to a scalar-tensor theory does not yield a dynamical equivalence between the two theories. This indicates that such Ostrogradsky-like reductions actually depend on the boundary conditions.\footnote{Intuitively, this is because boundary conditions determine whether the additional degrees of freedom in a theory can be organized in independent fields.} The mapping to the Einstein frame \cite{Amendola_Tsujikawa2010} of the scalar-tensor theory under the power-law condition is also addressed. The conformal transformation of the new boundary conditions yields a variational principle which is nonstandard and difficult to interpret physically; therefore, it is argued that the conformal mapping should also be confined to a particular set of boundary conditions.
\end{comment}

In section \ref{sec: Boundary variations in f(R) gravity}, the variation of the $f(R)$ action is reviewed, together with the usual choice $\delta R|_\mathcal{\partial M} = 0$ to obtain a simple GHY term. Although this choice can be mapped into the natural boundary conditions in the corresponding scalar-tensor theory, it is argued how this is not enough to prevent the search for generalized boundary conditions in $f(R)$ gravity. In subsection \ref{sec: Madsen-Barrow construction}, the construction of Madsen-Barrow GHY term is reviewed. Inspired by the GHY structure, in subsections \ref{sec:Power-law condition} and \ref{sec: Functional condition}, two more general boundary conditions on the scalar curvature are proposed, which encompass the usual case but enlarge the space of admissible variations. The consistency of these conditions with Gauss-Codazzi equation is discussed, and the expressions for the corresponding GHY terms are worked out explicitly. The perspectives for the physical implications of these results are discussed in the conclusions.

\section{Boundary variations in $f(R)$ gravity}
\label{sec: Boundary variations in f(R) gravity}
\begin{comment}
\textcolor{blue}{
\begin{itemize}
    \item[-] Variation of $f(R)$ action 
    \item[-] Boundary terms
    \item[-] Discussion: necessity of tightening boundary conditions (degrees of freedom): refs, no unique choice.
    \item[-] Discussion: why $\delta R = 0$ (typical choice) is more general than $\delta K = 0$: term $\propto \delta (n^\mu \nabla_\mu K)$ can be written as $K^2 + $ total variation (Wald). 
    \item[-] Discussion*: expression of $\delta R$ (here or appendix).   
\end{itemize}}
\end{comment}
The starting point of the discussion is the variation of the $f(R)$ action ($k = 8\pi G$)
\begin{equation}
    S_f = \frac{1}{2k}\int_\mathcal{M}d^Dx\sqrt{-g}f(R), 
    \label{eq2: f(R) action}
\end{equation}
where $f(R)$ is an arbitrary and non-linear function of the scalar curvature, while the factor $g$ is the determinant of the spacetime metric.\footnote{Throughout the paper, the Misner-Thorne-Wheeler “{+,+,+,+}" convention is adopted for the metric, the curvature tensors and the Einstein equations \cite{Misner2017}.} The integration is over the $D$-dimensional spacetime manifold $\mathcal{M}$ with boundary $\partial\mathcal{M}$.\footnote{The boundary is taken here to be a closed $(D-1)$-hypersurface (see Appendix \ref{app: Hypersurfaces}).} The metric is the only dynamical field in $S_f$; hence, the variations are constrained by a fixed configuration of the metric on the boundary $\delta g_{\mu\nu}|_\mathcal{\partial M} = 0$, which is assigned as a Dirichlet boundary condition for the variational principle. The variation gives (see Appendix \ref{app: Hypersurfaces})
\begin{multline}
\delta_g S_{f} = \frac{1}{2k}\int_\mathcal{M}d^Dx\sqrt{-g}\delta g_{\mu\nu}\left(\frac{1}{2}g^{\mu\nu} f(R) - \phi R^{\mu\nu} \right. \\\left. - g^{\mu\nu}\nabla^2f'(R) +  \nabla^\mu\nabla^\nu f'(R)\vphantom{\frac{1}{2}}\right) \\
-\frac{1}{k} \int_{\partial \mathcal{M}} d^dy \sqrt{h}\, \epsilon\, f'(R)\, \delta K,
\label{eq2: variation of the f(R) bulk action for outgoing normal vector}
\end{multline}
where the condition $\delta g_{\mu\nu}|_{\mathcal{\partial M}} = 0$ has already been imposed. 
The second term involves an integration over the $d = (D-1)$-dimensional boundary submanifold with proper coordinates $\{y^a\}$, induced metric $h_{ab}$ and outgoing normal vector $n_\mu$, which is assumed either timelike or spacelike, $n^\mu n_\mu = \epsilon \neq 0$.\footnote{The case of a lightlike normal vector is more involved and would need a separate treatment \cite{Poisson2009}.} The factor $\delta K$ is the variation of the trace of the boundary extrinsic curvature and it is proportional to the derivative of $g_{\mu\nu}$ across the boundary, Eq.\ (\ref{eqB: variation of the trace of the extrinsic curvature}); therefore, it is in general non vanishing even though $\delta g_{\mu\nu}|_{\mathcal{\partial M}} = 0$. 

Einstein's theory is recovered by setting $f'(R)\equiv 1$, in which case the boundary term in (\ref{eq2: variation of the f(R) bulk action for outgoing normal vector}) can be written as a total variation.\footnote{The embedding of the boundary surface in spacetime is considered, as usual, completely fixed in the variational principle; therefore, the normal vector and the induced metric are also fixed.} 
The GR gravitational action can thus be redefined by adding to the Einstein-Hilbert term the so-called Gibbons-Hawking-York boundary term \cite{York1972,Gibbons1977,Poisson2009}
\begin{equation}
    S_{GHY}^{GR} = \frac{1}{k}\int_{\partial \mathcal{M}} d^dy \sqrt{h}\, \epsilon\, K,
    \label{eq2: GHY term in GR}
\end{equation}
which does not affect the dynamical field equations but compensates the boundary variation deriving from the bulk part of the action. The complete gravitational action $S_G^{GR} = S_{EH} + S^{GR}_{GHY}$ is thus stationary on the solutions of the Einstein equations \cite{Poisson2009}.

In the more general $f(R)$ gravity case, there is consensus within contemporary literature in affirming that, without additional boundary conditions, the insertion of a non-constant $f'(R)$ in Eq.\ (\ref{eq2: variation of the f(R) bulk action for outgoing normal vector}) prevents the boundary term from being written as a total variation \cite{Dyer2008}. The physical motivation comes from the fact that the non-linearity of $f(R)$ is introducing an additional dynamical degree of freedom \cite{Belenchia2018higher}, which must be fixed on the boundary to allow well-posed variations. As in GR, fixing the extrinsic curvature of the boundary is a too restrictive condition on the space of solutions, as it would imply fixing the variation of the metric across the boundary. An attempt to construct a proper GHY term in $f(R)$ gravity without further restrictions on the variations at the boundary was made in \cite{Alhamawi2019}. A discussion on consistent boundary conditions of Dirichlet, Neumann and mixed type in $f(R)$ gravity, using its mapping to scalar-tensor theory \cite{Sotiriou2010,DeFelice2010,Capozziello2010}, together with the corresponding GHY terms, was conducted in \cite{Khodabakhshi2018, Khodabakhshi2020}.

The solution usually adopted in the literature \cite{Dyer2008} is to fix also the configuration of the scalar curvature on the boundary, $\delta R|_{\mathcal{\partial M}} = 0$. This is a weaker condition than fixing the extrinsic curvature, as can be deduced from the variation of Gauss-Codazzi equation (\ref{eqA: Gauss-Codazzi equation for the Ricci scalar}). Specifically, if $f''(R)$ is regular, the variation in the boundary term of Eq.\ (\ref{eq2: variation of the f(R) bulk action for outgoing normal vector}) can be moved out of the integral and the corresponding GHY immediately reads
\begin{equation}
    S_{GHY}^f|_{\delta R = 0} = \frac{1}{k} \int_{\partial \mathcal{M}} d^dy \sqrt{h}\, \epsilon\, f'(R)\,K,
    \label{eq2: GHY term in f(R) for delta R = 0}
\end{equation}
Naturally, for $f' = 1$ the GR case is recovered, Eq.\ (\ref{eq2: GHY term in GR}). As argued in \cite{Dyer2008}, the condition $\delta R|_{\mathcal{\partial M}} = 0$ with the corresponding GHY term (\ref{eq2: GHY term in f(R) for delta R = 0}) can be derived from the correspondence of $f(R)$ gravity with scalar-tensor theory \cite{Sotiriou2010,DeFelice2010,Capozziello2010}. Specifically, these are mapped from the natural conditions in scalar-tensor theory
\begin{equation}
    \delta \chi|_{\mathcal{\partial M}} = 0,\,\,\, S_{GHY}^{ST}|_{\delta \chi = 0} = \frac{1}{k} \int_{\partial \mathcal{M}} d^dy \sqrt{h}\, \epsilon\, f'(\chi)\,K,
    \label{eq2: natural boundary condition for the scalar field and corresponding GHY term in scalar-tensor theory}
\end{equation}
the field $\chi$ being the additional scalar. The authors of \cite{Dyer2008} then proceed to show that the expression of Schwarzschild black hole entropy, computed via Euclidean approach with the GHY term (\ref{eq2: GHY term in f(R) for delta R = 0}) in four dimensions, corresponds to the expression obtained via Wald's Noether-charge approach \cite{Wald1993black,Iyer1994some}.

These are strong motivations for adopting $\delta R|_{\mathcal{\partial M}} = 0$ as the additional boundary condition; however, encoding the degree of freedom arising from the nonlinearity of $f(R)$ in an independent scalar field - for which $\delta \chi|_{\mathcal{\partial M}} = 0$ is a natural boundary condition - is not by itself a physical requirement. Moreover, as explicitly acknowledged in \cite{Dyer2008}, Iyer and Wald have proven \cite{Iyer1995Comparison} that Noether charge and Euclidean approaches to black hole entropy are compatible in case of higher-derivative theories for which, among other conditions, the variational principle is well-posed with only the metric held fixed on the boundary. As argued above, this does not encompass the general $f(R)$ case. 

Such observations motivate the search for more general boundary conditions than fixing the scalar curvature, which provide nonetheless a well-posed stationary action principle in $f(R)$ gravity, and the study of their physical implications. In the next section, two more general conditions are proposed and the corresponding GHY boundary terms are computed. 

\section{Generalized GHY boundary terms}
\label{sec: More general Gibbons-Hawking-York boundary terms}
Before addressing the generalized GHY terms, the construction carried on in \cite{Madsen1989} is briefly reviewed. The resulting structure will be generalized and adapted to more general boundary conditions. 
\subsection{Madsen-Barrow construction}
\label{sec: Madsen-Barrow construction}
The starting point of \cite{Madsen1989} is the variation of Gauss-Codazzi equation (\ref{eqA: Gauss-Codazzi equation for the Ricci scalar})
\begin{multline}
    \delta R|_\mathcal{\partial M} =-\epsilon \left(2\frac{D+1}{D}K\delta K + 2\bar K^{\mu\nu} \delta\bar K_{\mu\nu}\right)\\ - 2\epsilon n^\mu\delta(\nabla_\mu K) + 2\epsilon\delta[\nabla_\alpha(n^\beta \nabla_\beta n^\alpha)],
    \label{eq3: variation of the D scalar curvature on the boundary with fixed induced metric}
\end{multline}
where the traceless part of the extrinsic curvature, $\bar K_{\mu\nu}$, has been separated. Notice that the variation is performed, as usual, with a completely fixed boundary, so that $\delta R^{D-1} = 0$. 

If the boundary is embedded in such a way that \cite{Madsen1989}
\begin{equation}
    \bar K^{\mu\nu} \delta\bar K_{\mu\nu} + n^\mu\delta(\nabla_\mu K) -\delta[\nabla_\alpha(n^\beta \nabla_\beta n^\alpha)] = 0,
    \label{eq3: Barrow-Madsen conditions on the variation of the extrinsic curvature on the boundary}
\end{equation}
which here implies
\begin{equation}
    \delta R|_\mathcal{\partial M} = -2\epsilon\frac{D+1}{D}K\delta K,
    \label{eq3: Madsen-Barrow boundary condition on the variation of the D-scalar curvature}
\end{equation}
then repeated integration by parts of the boundary term in Eq.\ (\ref{eq2: variation of the f(R) bulk action for outgoing normal vector}) yields the corresponding GHY term in the form of a convergent power series in the trace of the extrinsic curvature. In \cite{Madsen1989}, it is argued that the condition in Eq.\ (\ref{eq3: Barrow-Madsen conditions on the variation of the extrinsic curvature on the boundary}) is realized for boundaries of maximally symmetric spacetimes. Specifically, a first integration gives 
\begin{multline*}
    - \frac{1}{k} \int_{\partial\mathcal{M}}d^dy \sqrt{h}  \epsilon f'(R)\delta_g K =\\ - \frac{1}{k}\delta_g \int_{\partial\mathcal{M}}d^dy \sqrt{h}  \epsilon f'(R) K + \frac{1}{k} \int_{\partial\mathcal{M}}d^dy \sqrt{h}  \epsilon f''(R) K \delta_g R,
\end{multline*}
and insertion of the condition (\ref{eq3: Madsen-Barrow boundary condition on the variation of the D-scalar curvature}) leads to 
\begin{multline}
     - \frac{1}{k} \int_{\partial\mathcal{M}}d^dy \sqrt{h}  \epsilon f'(R)\delta_g K =\\ - \frac{1}{k}\delta_g \int_{\partial\mathcal{M}}d^dy\sqrt{h}\epsilon\left(Kf'(R) +\epsilon \frac{2+2D}{3D}K^3f''(R)\right) \\+\frac{1}{k}\frac{2+2D}{3D}\int_{\partial\mathcal{M}}d^dy\sqrt{h} K^3 f'''(R)\delta_gR.
\end{multline}
Iterating the procedure, it is straightforward to verify that the result is\footnote{Notice that Madsen and Barrow define the Riemann tensor with opposite sign with respect the convention adopted here; hence, the expression of the GHY term has opposite sign with respect to \cite{Madsen1989}.} \cite{Madsen1989}
\begin{multline}
    S^f_{GHY} = \frac{1}{k}\int_{\partial\mathcal{M}}d^dy\sqrt{h}\\\times\sum_{n = 1}^{\infty} \epsilon^n\left(\frac{2+2D}{D}\right)^{n-1} \frac{1}{(2n-1)!!}f^{(n)}(R)K^{2n-1}.
    \label{eq3: Madsen and Barrow GHY term}
\end{multline}
where $f^{(n)}(R)$ stands for the $n$-th derivative of the $f(R)$; hence, the $f$ needs to be differentiable to allow the construction. Notice that, if Eq.\ (\ref{eq3: Barrow-Madsen conditions on the variation of the extrinsic curvature on the boundary}) holds, the resulting condition (\ref{eq3: Madsen-Barrow boundary condition on the variation of the D-scalar curvature}) is more general than what is usually assumed, $\delta R|_\mathcal{\partial M} = 0$, the latter corresponding to the particular case in which $\delta K = 0$.

\subsection{Power-law condition}
\label{sec:Power-law condition}
\begin{comment}
\textcolor{blue}{
\begin{itemize}
    \item[-] Discussion: motivation: consider boundary condition which includes $\delta R = 0$ and Madsen-Barrow cases.
    \item[-] Introduce power law condition.  
    \item[-] Discussion: compatibility with Gauss-Codazzi --> Free boundary problems.
    \item[-] Discussion*: examples of free boundary (accreting black holes) 
    \item[-] GHY computation.
\end{itemize}}
\end{comment}

Motivated by Madsen and Barrow result, the following ansatz for the GHY term is considered
\begin{equation}
    2kS^f_{GHY} = -2\int_\mathcal{\partial M} d^dy \epsilon\sqrt{h}\sum_{n = 1}^{\infty} c_n f^{(n)}(R)K^{s_n},
    \label{eq3: general structure of GHY term for power law condition}
\end{equation}
where $s_n$ and $c_n$ are functions of the parameters characterizing the boundary conditions. These are determined by requiring that the variation of $S^f_{GHY}$ compensate the boundary variation in Eq.\ (\ref{eq2: variation of the f(R) bulk action for outgoing normal vector}).

In conjunction with (\ref{eq3: general structure of GHY term for power law condition}), a generalized power-law boundary condition for the scalar curvature is considered 
\begin{equation}
    \delta R|_{\partial \mathcal{M}} = \alpha K^r \delta K,\,\,\, \alpha = \alpha(y^a),r \in \mathbb{R},
    \label{eq3: power law condition for the variation of the scalar curvature on the boundary}
\end{equation}
with $\alpha$ a $C^1$ function of the intrinsic coordinates of the boundary, $\alpha: \mathbb{R}^3\to\mathbb{R}$.
This condition encompasses the usual case $\delta R|_{\mathcal{\partial M}} = 0$, for $\alpha \equiv 0$, and Madsen-Barrow condition for $\alpha \equiv -\epsilon\frac{2+2D}{D}$ and $r = 1$; therefore, it is a straightforward generalization. In general $c_n$ and $s_n$ will depend upon $\alpha$ and $r$.

The condition (\ref{eq3: power law condition for the variation of the scalar curvature on the boundary}) can be realized, for example, if the extrinsic curvature tensor has the form 
\begin{equation}
    K_{ab} = \frac{K}{D-1} h_{ab} + K^{\bar r} T_{ab},\,\,\, \bar r \in \mathbb{R}
    \label{eq3: structure of the extrinsic curvature to allow the power-law condition}
\end{equation}
with $T_{ab}$ any traceless tensor of the boundary submanifold, $h^{ab}T_{ab} =0$. In this case it is straightforward to verify that 
\begin{equation}
    K_{ab}K^{ab} = \frac{K^2}{D-1} + K^{2\bar r} T_{ab}T^{ab},
\end{equation}
so that Gauss-Codazzi equation becomes ($T_{ab}T^{ab} \equiv T^2$)
\begin{multline*}
    R|_\mathcal{\partial M} = R^{(D-1)} -\epsilon \left(\frac{D}{D-1}K^2 + T^2 K^{2\bar r}\right)\\ - 2\epsilon n^\mu \nabla_\mu K + 2\epsilon\nabla_\alpha(n^\beta \nabla_\beta n^\alpha).
\end{multline*}
If the intrinsic geometry of the boundary is fixed, the variation reads
\begin{multline*}
    \delta R|_\mathcal{\partial M} = - 2\epsilon\bar r T^2 K^{2\bar r-1}\delta K - 2\epsilon \frac{D}{D-1}K\delta K \\- 2\epsilon n^\mu \delta \nabla_\mu K + 2\epsilon\delta\nabla_\alpha(n^\beta \nabla_\beta n^\alpha).
\end{multline*}
Notice that the tensor $T_{ab}$ is not subject to variation; therefore, it must be provided \textit{a priori}, once the boundary submanifold structure is specified. The freedom in the choice of $T_{ab}$ essentially parametrizes the freedom in the choice of boundary conditions for the variational problem. The power-law condition (\ref{eq3: power law condition for the variation of the scalar curvature on the boundary}) is realized if the variations satisfy 
\begin{equation}
    \frac{D}{D-1} K\delta K + n^\mu \delta \nabla_\mu K - \delta\nabla_\alpha(n^\beta \nabla_\beta n^\alpha) = 0,
    \label{eq3: condition to realize power-law boundary condition from GC equation}
\end{equation}
and upon redefining $\alpha \equiv - 2\epsilon\bar r T^2$ and $r \equiv 2\bar r-1$. The constraint provided by Eq.\ (\ref{eq3: condition to realize power-law boundary condition from GC equation}) is the counterpart of Madsen-Barrow condition (\ref{eq3: Barrow-Madsen conditions on the variation of the extrinsic curvature on the boundary}); moreover, since the case (\ref{eq3: Madsen-Barrow boundary condition on the variation of the D-scalar curvature}) is encompassed by (\ref{eq3: power law condition for the variation of the scalar curvature on the boundary}), the constraint (\ref{eq3: condition to realize power-law boundary condition from GC equation}) is weaker with respect to (\ref{eq3: Barrow-Madsen conditions on the variation of the extrinsic curvature on the boundary}). As a last comment before addressing the variation of Eq.\ (\ref{eq3: general structure of GHY term for power law condition}), notice that, although the choice of $T_{ab}$ is arbitrary in principle, there can be natural choices convened by the symmetries of the boundary manifold. For example, if the boundary admits a normalized Killing vector field $\xi^a$, then a natural choice is $T_{ab} = \bar\alpha\left(\xi_a\xi_b - \frac{1}{D-1}h_{ab}\right)$, with $\bar \alpha$ a real constant.

The variation of the GHY term in Eq.\ (\ref{eq3: general structure of GHY term for power law condition}) reads
\begin{multline}
    2k\delta S^f_{GHY} = -2\int_\mathcal{\partial M} d^dy\epsilon \sqrt{h} \delta K\sum_{n = 1}^{\infty} c_n\left[\alpha f^{(n+1)}(R)K^{s_n+r}\right. \\\left.+ s_nf^{(n)}(R)K^{s_n-1}\right],
    \label{eq3: variation of the GHY ansatz for power law condition}
\end{multline}
while asking that it matches the boundary term in Eq.\ (\ref{eq2: variation of the f(R) bulk action for outgoing normal vector}) provides the equation 
\begin{multline}
    c_1 f'(R) s_1K^{s_1-1} +...+f^{(l)}(R)\left[c_{l-1}\alpha K^{s_{l-1}+r} + c_l s_lK^{s_l-1}\right] \\+ ... = -  f'(R).
    \label{eq3: matching condition for power law}
\end{multline}
A possibility to satisfy the identity is by taking 
\begin{equation}
    s_1 \equiv 1,\,\, c_1 = -1,\,\, c_l = -\alpha\frac{c_{l-1}}{s_l} K^{s_{l-1}+ r - s_l+1},
    \label{eq: matching of the coefficients of the derivatives of f in the power law ghy}
\end{equation}
which also implies
\begin{equation}
    s_l = s_{l-1} + r+ 1, \,\, s_1 = 1,
\end{equation}
since the $c_l$'s cannot depend upon $K$. Notice that Eq.\ (\ref{eq: matching of the coefficients of the derivatives of f in the power law ghy}) is a sufficient condition, unless the derivatives of the $f(R)$ are all independent. However, this appears to be the only possibility to build explicitly a GHY term without any assumption on the $f(R)$ form. The recursive equation for $s_l$ is solved by taking
\begin{equation}
    s_l = l+(l-1)r. 
    \label{eq3: recursive relation for the exponents in the GHY ansatz of power law}
\end{equation}
If $r = -1$, then $s_l = 1,\,\, \forall l$ and $c_n = (-1)^n\alpha^{n-1}$. The corresponding GHY term reads
\begin{equation}
\begin{aligned}
    &2kS^f_{GHY} = -2\int_\mathcal{\partial M} d^dy\epsilon \sqrt{h} K\sum_{n = 1}^{\infty} (-1)^n\alpha^{n-1} f^{(n)}(R),\\
    &\delta R|_{\mathcal{\partial M}} = \alpha \frac{\delta K}{K}.
\end{aligned}
    \label{eq3: power-law GHY term with r = -1}
\end{equation}
If $r \neq -1$, the $c_l$'s satisfy 
\begin{equation}
    c_l = -\frac{\alpha}{1+r} \frac{1}{\frac{1}{1+r} + l -1} c_{l-1},\,\,\, c_1 = -1.
\end{equation}
The solution is straightforward and reads\footnote{This can be obtained by writing the result, first, in terms of the raising Pochhammer symbol, $a^{[n]}\equiv a(a+1)...(a+n-1)$, and then recalling that $a^{[n]} =\frac{\Gamma(a+n)}{\Gamma(a)}$.}
\begin{equation}
    c_n = (-1)^n\frac{\alpha^{n-1}}{(1+r)^n}\frac{\Gamma\left(\frac{1}{1+r}\right)}{\Gamma\left(\frac{1}{1+r} + n\right)},
\end{equation}
where $\Gamma$ is the usual Euler $\Gamma$-function. Notice that, for $r = 1$, the $c_n$'s reduce to $c_n = \frac{(-1)^n\alpha^{n-1}}{(2n-1)!!}$, in agreement with Madsen and Barrow case \cite{Madsen1989} for $\alpha = -\epsilon\frac{2+2D}{D}$. Therefore, the GHY term corresponding to the condition (\ref{eq3: power law condition for the variation of the scalar curvature on the boundary}) is 
\begin{multline}
    S^f_{GHY} = \frac{1}{k}\int d^dy \sqrt{h}\epsilon f'(R)K -\frac{1}{k}\int d^dy \sqrt{h}\epsilon\\ \times\sum_{n = 2}^{\infty} (-1)^{n} \frac{\alpha^{n-1}}{(1+r)^{n}} \frac{\Gamma\left(\frac{1}{1+r}\right)}{\Gamma\left(\frac{1+(1+r)n}{1+r}\right)} f^{(n)}(R)K^{n+(n-1)r}.
    \label{eq3: GHY term for power-law condition}
\end{multline}
Notice how the usual case of Eq.\ (\ref{eq2: GHY term in f(R) for delta R = 0}) is recovered for $\alpha = 0$.

Also, notice how the derivation from Eq.\ (\ref{eq3: variation of the GHY ansatz for power law condition}) holds as long as all the $s_n$'s are non-vanishing. If for some $l^\star$ the $s_{l^\star+1}$
were vanishing, from Eq.\ (\ref{eq: matching of the coefficients of the derivatives of f in the power law ghy}) the coefficient $c_{l^\star}$ should also vanish, together with the descending tower of coefficients down to $c_1$, and the matching condition (\ref{eq3: matching condition for power law}) could not be satisfied without additional assumptions on the $f$. From Eq.\ (\ref{eq3: recursive relation for the exponents in the GHY ansatz of power law}), it is clear that if $\frac{1}{1+r}$ is a negative integer there will be an $l^\star$ so that $s_{l^\star+1} = 0$. This means that the boundary conditions in which $r = -1 - \frac{1}{m}$, should be discarded.  

The convergence of the series in Eq.\ (\ref{eq3: GHY term for power-law condition}) is discussed in Appendix \ref{sec: Convergence of generalized GHY series}; in the next section a more general functional condition is considered.

\subsection{Functional condition}
\label{sec: Functional condition}
\begin{comment}
\textcolor{blue}{\begin{itemize}
    \item[-] Discussion: motivation: even more general condition. 
    \item[-] GHY computation.
\end{itemize}}
\end{comment}
Generalizing the previous case, in this section it is considered the condition 
\begin{equation}
    \delta R|_{\partial M} = g(K)\delta K, 
    \label{eq3: general function of the extrinsic curvature condition for the variation of the scalar curvature on the boundary}
\end{equation}
where $g$ is kept an arbitrary function for the moment, in conjunction with the ansatz for the GHY term
\begin{equation}
2k    S^f_{GHY} = -2\int_\mathcal{\partial M} d^dy \sqrt{h}\epsilon \sum_{n = 1}^{\infty} f^{(n)}(R)G(K,n), 
    \label{eq3: ansatz for the GHY structure corresponding to the functional condition}
\end{equation}
where $G(K,n)$ is a function which, as before, is determined by the condition of matching the boundary term in Eq.\ (\ref{eq2: variation of the f(R) bulk action for outgoing normal vector}). The condition (\ref{eq3: general function of the extrinsic curvature condition for the variation of the scalar curvature on the boundary}) can be obtained, as in the previous section, by considering the extrinsic curvature of the form 
\begin{equation}
    K_{ab} = \frac{K}{D-1}h_{ab} + \bar g(K)T_{ab},
    \label{eq3: structure of the extrinsic curvature of the boundary to allow the functional condition}
\end{equation}
with $T_{ab}$ traceless and $\bar g$ at least $C^1$, and by imposing the constraint (\ref{eq3: condition to realize power-law boundary condition from GC equation}). 

After variation, the condition reads 
\begin{multline}
    f'(R)\partial_K G(K,1) + ... + f^{(l)}(R)\left[G(K,l-1)g(K) + \partial_K G(K,l)\right]\\ + ... = - f'(R). 
\end{multline}
Without any assumption on the form of the $f(R)$, a sufficient condition is
\begin{equation}
    \partial_K G(K,1) = -1,\,\,\, \frac{G(K,l-1)}{\partial_K G(K,l)}g(K) = -1.
    \label{eq3: conditions on the coefficients of the Barrow-type boundary term for a g(K) condition}
\end{equation}
The first equation gives $G(K,1) = - K$, where the integration constant is set by asking that the usual term (\ref{eq2: GHY term in f(R) for delta R = 0}) is recovered when $g(K)\equiv 0$. To solve the second equation, the function $g(K)$ is assumed everywhere analytic, so that it can be written as a convergent power series
\begin{equation}
    g(K) = \sum_{s= 0}^{+\infty} \frac{\beta_s}{s!}K^s.
    \label{eq3: Taylor expansion of g(K)}
\end{equation}
Equation (\ref{eq3: conditions on the coefficients of the Barrow-type boundary term for a g(K) condition}) can now be addressed recursively. For example, for the term $G(K,2)$ 
\begin{multline*}
\partial_K G(K,2) = \sum_{s = 0}^{+\infty} \frac{\beta_s}{s!}K^{s+1} \\\implies G(K,2) = \sum_{s = 0}^{+\infty}\frac{\beta_s}{(s+2)s!}K^{s+2},
\end{multline*}
while for the term $G(K,3)$
\begin{multline*}
     \partial_K G(K,3)  = - \sum_{s,t} \frac{\beta_s\beta_t}{(s+2)s!t!}K^{s+t+2} \\ \implies G(K,3) = - \sum_{s,t} \frac{\beta_s\beta_t}{(s+2)s!(s+t+3)t!}K^{s+t+3}.
\end{multline*}
By generalizing this structure, the expression for $G(K,l)$ is found $(l>1)$
\begin{equation}
    G(K,l) = (-1)^l\sum\limits_{\{s_i\}_{i = 1...l-1}}\frac{\left(\prod\limits_{i= 1}^{l-1}\beta_{s_i}\right)K^{\sum\limits_{i = 1}^{l-1}s_i + l}}{\left(\prod\limits_{i =1}^{l-1}s_i!\right) \prod\limits_{i = 1}^{l-1}\left(\sum\limits_{j\leq i}s_j + i + 1\right)}. 
\end{equation}
The expression for the GHY term corresponding to the boundary conditions (\ref{eq3: general function of the extrinsic curvature condition for the variation of the scalar curvature on the boundary}) and (\ref{eq3: Taylor expansion of g(K)}) follows 
\begin{multline}
        S^f_{GHY} = \frac{1}{k}\int d^dy \sqrt{h}\epsilon f'(R)K  
       \\ - \frac{1}{k}\int d^dy \sqrt{h}\epsilon  \sum\limits_{n = 2}^{\infty}(-1)^nf^{(n)}(R)\\\times \sum\limits_{\{s_i\}_{i = 1...n-1}}\frac{\left(\prod\limits_{i= 1}^{n-1}\beta_{s_i}\right)K^{\sum\limits_{i = 1}^{n-1}s_i + n}}{\left(\prod\limits_{i =1}^{n-1}s_i!\right) \prod\limits_{i = 1}^{n-1}\left(\sum\limits_{j\leq i}s_j + i + 1\right)}.
    \label{eq3: GHY term corresponding to the general functional boundary condition for the scalar curvature}
\end{multline}
The perspectives for the physical applications of these results are discussed in the conclusions.

\section{Conclusions}
\begin{comment}
\textcolor{blue}{
\begin{itemize}
    \item[-] The boundary conditions cannot change the dofs in a theory, but it can change the way in which the dofs are distributed among the fields (hot water), as well as the number of dynamical fields. The physics can be more or less clear with a given choice of conditions, but this is not a necessary condition. Notice also that every choice of bc changes the space of admissible solutions.   
    \item[-] Some perspectives (brief) on BH thermo.
\end{itemize}}
\end{comment}
In any classical field theory, it is important to allow the widest possible space of variations, so that physical solutions are not excluded \textit{a priori}. The choice of boundary conditions on the dynamical fields becomes crucial, especially when no natural or physically motivated choices are available. 

In this paper, the problem was addressed in the context of $f(R)$ gravity. In the preliminary section \ref{sec: Boundary variations in f(R) gravity} it was shown that, even in the free gravity case and contrary to standard GR, fixing only the metric on the boundary is no more sufficient to ensure a well-posed stationary action principle. Moreover, it was argued how this problem is commonly solved by further providing the configuration of the scalar curvature on the boundary, $\delta R|_\mathcal{\partial M} = 0$. Although this choice allows a particularly simple expression for the GHY term and it is easily interpretable in terms of the $f(R)$ to scalar-tensor mapping \cite{Dyer2008}, it is nonetheless important to ask whether more general or alternative choices exist and allow a well-posed action principle. In this direction, a first investigation was conducted by Madsen and Barrow \cite{Madsen1989}, who indeed found an alternative structure for the GHY term under the condition $\delta R|_\mathcal{\partial M} \propto K\delta K$.

Inspired by this GHY construction, in subsection \ref{sec:Power-law condition} a more general power-law boundary condition on the variation of the scalar curvature was proposed, Eq.\ (\ref{eq3: power law condition for the variation of the scalar curvature on the boundary}), in conjunction with the ansatz (\ref{eq3: general structure of GHY term for power law condition}) for the structure of the corresponding GHY term. Such boundary condition encompasses both the usual case and Madsen-Barrow condition and can be obtained from the variation of Gauss-Codazzi equation, if the boundary satisfies Eqs.\ (\ref{eq3: structure of the extrinsic curvature to allow the power-law condition}) and (\ref{eq3: condition to realize power-law boundary condition from GC equation}). The reconstruction of the actual form of the corresponding GHY term was worked out in details, confirming that it can be written as a convergent power-series in $K$, Eq.\ (\ref{eq3: GHY term for power-law condition}).

In subsection \ref{sec: Functional condition} the condition on $\delta R|_{\mathcal{\partial M}}$ was further generalized to a functional condition, Eqs.\ (\ref{eq3: general function of the extrinsic curvature condition for the variation of the scalar curvature on the boundary}) and (\ref{eq3: Taylor expansion of g(K)}), with a similar ansatz for the GHY term, Eq.\ (\ref{eq3: ansatz for the GHY structure corresponding to the functional condition}). As before, the functional condition can be obtained from Gauss-Codazzi equation if the extrinsic geometry of the boundary satisfies Eqs.\ (\ref{eq3: structure of the extrinsic curvature of the boundary to allow the functional condition})  and (\ref{eq3: condition to realize power-law boundary condition from GC equation}). The expression of the GHY term was again explicitly found as a more complicated power series in $K$, Eq.\ (\ref{eq3: GHY term corresponding to the general functional boundary condition for the scalar curvature}).

The expressions (\ref{eq3: GHY term for power-law condition}) and (\ref{eq3: GHY term corresponding to the general functional boundary condition for the scalar curvature}), together with the corresponding boundary conditions (\ref{eq3: power law condition for the variation of the scalar curvature on the boundary}) and (\ref{eq3: general function of the extrinsic curvature condition for the variation of the scalar curvature on the boundary}), in principle widen the space of solutions over which the action principle is well-posed in $f(R)$ gravity. Strictly speaking, the larger space of solutions allowed by, say, Eq.\ (\ref{eq3: power law condition for the variation of the scalar curvature on the boundary}) could be empty for $\alpha \neq 0, -2\epsilon \frac{D+1}{D}$. It would be important, therefore, to find an actual bulk solution lying in this larger space, along with a boundary surface satisfying Eqs.\ (\ref{eq3: structure of the extrinsic curvature to allow the power-law condition}) and (\ref{eq3: condition to realize power-law boundary condition from GC equation}). This is potentially a very hard task, which will be subject of forthcoming research. It should be mentioned, however, that the constructions made so far are nonetheless useful as formal generalisations that encompass the cases already known in the literature.

Finally, it is worth mentioning two intriguing research directions stemming from the direct application of the results presented here. The first, more formal, has to do with the mapping of $f(R)$ gravity into a scalar-tensor theory. As argued in section \ref{sec: Boundary variations in f(R) gravity} and in \cite{Dyer2008}, the usual condition $\delta R|_{\mathcal{\partial M}} = 0$ and the corresponding GHY term (\ref{eq2: GHY term in f(R) for delta R = 0}) emerge from the mapping of the natural choices (\ref{eq2: natural boundary condition for the scalar field and corresponding GHY term in scalar-tensor theory}) in scalar-tensor theory. Reversing the argument, it would be interesting to understand how the generalized boundary conditions proposed here, and corresponding GHY terms, are mapped and interpreted in scalar-tensor theory, and whether the mapping itself is affected. 

The second direction is somewhat more closely related to physical applications. As also pointed out by Dyer and Hinterbichler \cite{Dyer2008}, boundary terms in the $f(R)$ action generally determine the expressions of the ADM energy and of the entropy of a black hole, particularly when the latter is computed via the Euclidean approach. Therefore, the new GHY terms computed here can provide generalized expressions for these quantities and the laws of black-hole thermodynamics, carrying new physical insights. This last point will be addressed in forthcoming work \cite{Scali2026BHthermo}.
\begin{acknowledgments}
F.\ S.\ and S.\ L.\ C.\ wish to thank G.\ Bazzoni for a preliminary discussion on the mathematical properties of deformable boundaries. F.\ S.\ also thanks L.\ Amendola, J.\ Fabris, S.\ Liberati, O.\ Luongo and O.\ Zanusso for the chance to present preliminary versions of this work and useful discussions. F.\ S.\ is especially grateful to O.\ F.\ Piattella and A.\ Lapi for mentorship and invaluable help. F.\ S.\ further thanks the Insubria Theory Group for many spontaneous exchanges. 
\end{acknowledgments}
\bibliographystyle{ieeetr}
\bibliography{Bibliography.bib}

\appendix 

\section{Hypersurfaces}
\label{app: Hypersurfaces}
\begin{comment}
\textcolor{blue}{
\begin{itemize}
    \item[-] Expression of $\delta R = \delta R^{(d)}+...$ (Gauss-Codazzi)
    \item[-] Discussion: Wald E.2.28 and Madsen 35 and the role of total derivative ($\nabla_\mu(n^\mu K)$) in GR and $f(R)$ (can they be discarded also in $f(R)$?)
    \end{itemize}}
\end{comment}
In this section, the basic notions concerning the theory of embedded hypersurfaces are reviewed \cite{Poisson2009,Nakahara2003}. 

Let $\mathcal{M}$ be the $D$-dimensional pseudo-Riemannian spacetime manifold endowed with a Lorentzian metric $g$ with mostly plus signature. Let $\mathcal{N}$ be a $(D-1)$-dimensional Riemannian manifold, isometrically embedded in spacetime by means of the differentiable map $\phi: \mathcal{N} \to \mathcal{M}$. The image $\Sigma = \phi(\mathcal{N})$ is a hypersurface in spacetime, which is taken to be nowhere null. The coordinates of open sets in $\Sigma$ can be expressed as functions $\{x^\mu(y^a)\}$, $\mu = 0...D-1$, of the proper coordinates $\{y^a\}$, $a = 1...D-1$, in $\mathcal{N}$. 

At any point $p \in \mathcal{N}$, the embedding map defines an injection $\phi_\star : T_p(\mathcal{N}) \to T_{\phi(p)}\mathcal{M}$ between the tangent spaces at $p$ and $\phi(p)$. A vector $v_\mathcal{N} \in T_p(\mathcal{N})$ can thus be \textit{pushed forward} to 
\begin{equation}
    v_\mathcal{M} = \phi_\star v_\mathcal{N}, \text{\,\,\,\,or in coordinates,\,\,\,\,}  v_\mathcal{M}^\mu = \pd{x^\mu}{y^a} v_\mathcal{N}^a.
\end{equation}
Analogously, the embedding defines an injection $\phi^\star$ of the dual spaces $T^\star_p(\mathcal{N})$ and $T^\star_{\phi(p)}(\mathcal{M})$, so that a $1$-form can be \textit{pulled back} 
\begin{equation}
    \omega_\mathcal{N}  = \phi^\star \omega_\mathcal{M}, \text{\,\,\,\,or in coordinates,\,\,\,\,} (\omega_\mathcal{M})_a = \pd{x^\mu}{y^a}(\omega_\mathcal{N})_\mu.
\end{equation}
The extension to $(q,0)$ and $(0,p)$ tensors is straightforward \cite{Nakahara2003}.

The unit normal vector field $n$ to $\Sigma$ in spacetime is defined as the vector satisfying
\begin{equation}
    n_\mu \pd{x^\mu}{y^a} = 0,\,\,\, n^\mu n_\mu = \epsilon,
\end{equation}
where $\epsilon = -1 (+1)$ if $\Sigma$ is spacelike (timelike).  A tensor satisfying 
\begin{equation}
    n_\alpha T^{\alpha\beta...} = n_\beta T^{\alpha\beta...} = ... = 0,
\end{equation}
is tangent to $\Sigma$ and can be decomposed as
\begin{equation}
    T^{\alpha\beta...} = \pd{x^\alpha}{y^a}\pd{x^\beta}{y^b}...T^{ab...}.
\end{equation}
The projector onto $T(\Sigma)$ can be defined as
\begin{equation}
    h^\mu_\nu \equiv \delta ^\mu_\nu - \epsilon n^\mu n_\nu,
    \label{eq: tangential projector to the hypersurface}
\end{equation}
so that, given a tensor $Q \in T(\mathcal{M})$, the tensor 
\begin{equation}
    Q_{||}^{\alpha_1\alpha_2...} = h^{\alpha_1}_{\beta_1}h^{\alpha_2}_{\beta_2}... Q^{\beta_1\beta_2...},
\end{equation}
is tangent to $\Sigma$. 

Naturally, the embedding map $\phi$ induces a metric (or first fundamental form) on $\mathcal{N}$ via pull back of the spacetime metric
\begin{equation}
    h = \phi^\star g, \text{\,\,\,\,or in coordinates,\,\,\,\,} h_{ab} = \pd{x^\mu}{y^a}\pd{x^\nu}{y^b} g_{\mu\nu}.
\end{equation}
The covariant derivative in $\mathcal{N}$ is defined by projection of spacetime metric-compatible covariant derivative
\begin{equation}
D_a A_b  \equiv \pd{x^\mu}{y^a}\pd{x^\nu}{y^b} \nabla_\mu A_\nu.
\label{eq: intrinsic covariant derivative definition}
\end{equation}
The latter can be written as 
\begin{equation}
    D_a A_b = \partial_a A_b - \Gamma^c_{ab}A_c,
\end{equation}
where $\Gamma^c_{ab}$ are coefficients of the connection compatible with the induced metric 
\begin{equation}
\Gamma^a_{bc} = \frac{1}{2} h^{ad} \{ \partial_b h_{dc} + \partial_c h_{db}  -
 \partial_d h_{bc}  \}.
 \label{eq: Induced Connection Coefficients}
\end{equation}
The induced metric retains, in some way, all the information about the intrinsic geometric properties of $\Sigma$. The extrinsic curvature of the surface (or second fundamental form) is instead defined as 
\begin{equation}
K_{\alpha\beta} =  h^{\mu}_{\alpha}h^{\nu}_{\beta}\nabla_\mu n_\nu. 
\label{eq: Second Fundamental Form Definition}
\end{equation}
The latter is symmetric and tangent with respect to $\Sigma$, and can be equivalently defined as a tensor on $\mathcal{N}$  
\begin{equation}
    K_{ab} =  \pd{x^\mu}{y^a}\pd{x^\nu}{y^b}\nabla_\mu n_\nu. 
\end{equation}

Finally, the intrinsic curvature of $\mathcal{N}$ is naturally defined as
\begin{equation}
(R^{(D-1)})\indices{^a_{bcd}} = \partial_c \Gamma^a_{bd} - \partial_d \Gamma^a_{bc} + \Gamma^a_{ce} \Gamma^e_{bd} - \Gamma^a_{de} \Gamma^e_{cd},
\label{eq: Riemann 3-tensor expression}
\end{equation}
which is related to the spacetime curvature by Gauss-Codazzi equation 
\begin{equation}
R^{(D-1)}_{abcd} = \pd{x^\alpha}{y^a}\pd{x^\beta}{y^b}\pd{x^\gamma}{y^c}\pd{x^\delta}{y^d} R_{\alpha\beta\gamma\delta} 
-\epsilon (K_{ad}K_{bc} - K_{ac}K_{bd}).
\label{eq: Generalized Gauss-Codazzi}
\end{equation}
Contraction of the intrinsic Riemann tensor by means of the induced metric provides the intrinsic Ricci scalar. It is useful to write the result as the expression of the spacetime Ricci scalar evaluated on $\Sigma$ 
\begin{multline}
    R|_\Sigma = R^{(D-1)} -\epsilon \left(K^2 + K_{\mu\nu}K^{\mu\nu}\right)\\ - 2\epsilon n^\mu \nabla_\mu K + 2\epsilon\nabla_\alpha(n^\beta \nabla_\beta n^\alpha),
    \label{eqA: Gauss-Codazzi equation for the Ricci scalar}
\end{multline}
in terms of the intrinsic Ricci scalar.

\section{Variation of {$f(R)$} action}
In this section, the variation of the $f(R)$ action (\ref{eq2: f(R) action}) is detailed \cite{Capozziello2010,Scali2024} in a spacetime region with boundary $\partial \mathcal{M}$. The variations are constrained by $\delta g_{\mu\nu}|_{\partial \mathcal{M}} = 0$ and a fixed boundary, so that $\delta h_{ab} = \delta n_\mu = 0$. 

Specifically, the variation gives
\begin{equation}
\begin{aligned}
    \delta_g S_{f} =& \frac{1}{2k}\int_\mathcal{M}d^Dx\sqrt{-g}\delta g_{\mu\nu}\left(\frac{1}{2}g^{\mu\nu} f(R) - f'(R) R^{\mu\nu}\right) \\& - 
    \frac{1}{2k}\int_\mathcal{M} d^Dx\sqrt{-g}f'(R)(g^{\rho\sigma}g^{\mu\nu} - g^{\rho\mu}g^{\sigma\nu})\nabla_\rho \nabla_\sigma \delta g_{\mu\nu}.
\end{aligned}
\end{equation}
Integrating by parts once, the second line can be split in a bulk plus a boundary part
\begin{multline}
    - 
    \frac{1}{2k}\int_\mathcal{M} d^Dx\sqrt{-g}f'(R)(g^{\rho\sigma}g^{\mu\nu} - g^{\rho\mu}g^{\sigma\nu})\nabla_\rho \nabla_\sigma \delta g_{\mu\nu} =\\
    -\frac{1}{2k}\int_\mathcal{\partial M} d^dy\sqrt{h}\,\epsilon\, n_\rho[f'(R)(g^{\rho\sigma}g^{\mu\nu} - g^{\rho\mu}g^{\sigma\nu}) \partial_\sigma \delta g_{\mu\nu}]
      \\+
    \frac{1}{2k}\int_\mathcal{M} d^Dx\sqrt{-g}
    (\nabla_\rho f'(R))(g^{\rho\sigma}g^{\mu\nu} - g^{\rho\mu}g^{\sigma\nu}) \nabla_\sigma \delta g_{\mu\nu},
\label{eq: splitting of the term coming from the variation of the Ricci tensor in bulk plus boundary parts}
\end{multline}
where $\alpha$ is a number taking values $+1$ for outgoing $n$ and $-1$ for ingoing $n$. Notice that $\nabla_{\sigma} \delta g_{\mu\nu}|_{\partial\mathcal{M}} = \partial_\sigma \delta g_{\mu\nu} |_{\partial\mathcal{M}}$ because of the boundary condition. By expanding the metric by means of the tangential projector in Eq.\ (\ref{eq: tangential projector to the hypersurface}), that is, $g_{\mu\nu}|_\mathcal{\partial M} = h_{\mu\nu} + \epsilon n_\mu n_\nu$, the boundary term can be written as 
\begin{multline}
    -\frac{1}{2k}\int_\mathcal{\partial M} d^dy\sqrt{h}\,\epsilon\, n_\rho[f'(R)(g^{\rho\sigma}g^{\mu\nu} - g^{\rho\mu}g^{\sigma\nu}) \partial_\sigma \delta g_{\mu\nu}] =\\ -\frac{1}{2k} \int_{\partial \mathcal{M}} d^dy \sqrt{h}\, \epsilon\, f'(R)\, h^{\mu\nu}n^\sigma \partial_\sigma\delta g_{\mu\nu} \\+ \frac{1}{2k}\int_{\mathcal{\partial M}} d^dy \sqrt{h}\, \epsilon\,f'(R)\, h^{\mu\beta}n^\alpha \partial_\mu \delta g_{\alpha \beta}.
    \label{eq2: boundary terms from the varition of the f(R) action}
\end{multline}
Since $\delta g_{\mu\nu}$ is identically zero on the boundary, the tangential derivative $h^{\mu\beta}\partial_\mu \delta g_{\alpha \beta}$ vanishes and so is the corresponding term in Eq.\ (\ref{eq2: boundary terms from the varition of the f(R) action}). By the same token, the derivative across the boundary, $n^\sigma \partial_\sigma\delta g_{\mu\nu}$, is not fixed by the boundary condition, so that the second line in Eq.\ (\ref{eq2: boundary terms from the varition of the f(R) action}) does not vanish. 

The bulk term in Eq.\ (\ref{eq: splitting of the term coming from the variation of the Ricci tensor in bulk plus boundary parts}) can again be integrated by parts, neglecting this time boundary contributions since no derivative is left to act on $\delta g_{\mu\nu}$. Putting everything together, the variation of the $f(R)$ action reads
\begin{multline}
\delta_g S_{f} = \frac{1}{2k}\int_\mathcal{M}d^Dx\sqrt{-g}\delta g_{\mu\nu}\left(\frac{1}{2}g^{\mu\nu} f(R) - \phi R^{\mu\nu} \right. \\\left. - g^{\mu\nu}\nabla^2f'(R) +  \nabla^\mu\nabla^\nu f'(R)\vphantom{\frac{1}{2}}\right) \\
-\frac{1}{2k} \int_{\partial \mathcal{M}} d^dy \sqrt{h}\, \epsilon\, f'(R)\, h^{\mu\nu}n^\sigma \partial_\sigma\delta g_{\mu\nu}.
\end{multline}
As it happens in GR, the variation of the trace $K$ of the extrinsic curvature can be used to simplify this expression \cite{Poisson2009}
\begin{equation}
    \delta K = \delta (\nabla_\mu n^\mu) = -h^{\mu\nu}\delta \Gamma^\rho_{\mu\nu} n_\rho = \frac{1}{2}n^\rho h^{\mu\nu}  \partial_\rho \delta g_{\mu\nu}.
    \label{eqB: variation of the trace of the extrinsic curvature}
\end{equation}
It follows 
\begin{multline}
\delta_g S_{f} = \frac{1}{2k}\int_\mathcal{M}d^Dx\sqrt{-g}\delta g_{\mu\nu}\left(\frac{1}{2}g^{\mu\nu} f(R) - \phi R^{\mu\nu} \right. \\\left. - g^{\mu\nu}\nabla^2f'(R) +  \nabla^\mu\nabla^\nu f'(R)\vphantom{\frac{1}{2}}\right) \\
-\frac{1}{k} \int_{\partial \mathcal{M}} d^dy \sqrt{h}\, \epsilon\, f'(R)\, \delta K,
\end{multline}
for the total variation of the $f(R)$ action.

\section{Convergence of generalized GHY series}
\label{sec: Convergence of generalized GHY series}
In this section, the convergence of the series in the power-law GHY term (\ref{eq3: GHY term for power-law condition}) is discussed. For a general complex power series $\sum_n a_n z^n$ the radius of convergence $Z$ is given by 
\begin{equation}
    \frac 1Z=\limsup_{n\to\infty}|a_n|^{\frac 1n}. 
\end{equation}
Therefore, for the series in Eq.\ (\ref{eq3: GHY term for power-law condition}), the radius of convergence is given by 
\begin{align}
    \frac 1Z=&\limsup_{n\to\infty}\left(\frac{\alpha^{n-1}}{(1+r)^{n}} \frac{1}{\Gamma\left(\frac{1+(1+r)n}{1+r}\right)} |f^{(n)}(R)|\right)^{\frac 1{n|r+1|}}\cr
    =&\left( \frac \alpha{r+1} \right)^{\frac 1{|r+1|}} \limsup_{n\to\infty}\left( \frac{1}{\Gamma\left(\frac{1+(1+r)n}{1+r}\right)} |f^{(n)}(R)|\right)^{\frac 1{n|r+1|}}.
\end{align}
Since $f(z)$\footnote{For generality, here it is considered the analytic continuation of the $f$ to complex arguments.} is assumed analytic in $z = R$, the minimum distance between $R$ and the nearest singularity of $f$ is given by 
\begin{equation}
    \frac 1d= \limsup_{n\to\infty}\left|\frac {f^{(n)}(R)}{n!} \right|^{\frac 1n}.
\end{equation}
The expression of $Z$ simplifies to 
\begin{equation}
    \frac{1}{Z}=\left( \frac \alpha{(r+1)d} \right)^{\frac 1{|r+1|}} \lim_{n\to\infty}\left( \frac{n!}{\Gamma\left(\frac{1}{1+r}+n\right)} \right)^{\frac 1{n|r+1|}}.
    \label{eqC: residual n limit in convergence radius definition}
\end{equation}
By employing $\Gamma(n+1) = n!$ and that 
\begin{equation}
    \Gamma\left(n + \frac{1}{1+r}\right) \simeq \Gamma(n)n^\frac{1}{1+r} = \frac{n!}{n^{\frac{r}{1+r}}},
\end{equation}
for large $n$, the remaining limit in Eq.\ (\ref{eqC: residual n limit in convergence radius definition}) tends to unity and $Z$ is given by
\begin{equation}
    Z=\left( \frac {(r+1)d}\alpha \right)^{\frac 1{|r+1|}}.
\end{equation}
Therefore, the calculations in section \ref{sec:Power-law condition} are automatically justified as long as $|K| \leq Z$, for which the series in (\ref{eq3: GHY term for power-law condition}) converges. Of course, if $f(z)$ is everywhere analytic in $\mathbb{C}$, then the convergence radius is infinite and there is no restriction on the values of $K$.
\end{document}